\documentclass[aps,pra,reprint,twocolumn,amsmath,amssymb,citeautoscript]{revtex4-1}
\usepackage{graphicx}
\usepackage{dcolumn}
\usepackage{bm}
\usepackage{latexsym,epsfig}
\usepackage{graphicx}
\usepackage{verbatim}
\usepackage{comment}
\usepackage{amsmath}
\usepackage{amssymb}
\usepackage{stmaryrd}
\usepackage{color}
\usepackage{epstopdf}
\usepackage{grffile}
\usepackage{float}
\DeclareGraphicsExtensions{.eps}

\usepackage{ulem}
\newcommand{\beq}{\begin{equation}}
\newcommand{\eeq}{\end{equation}}
\newcommand{\bea}{\begin{eqnarray}}
\newcommand{\eea}{\end{eqnarray}}
\newcommand{\ben}{\begin{eqnarray*}}
\newcommand{\een}{\end{eqnarray*}}
\newcommand{\bfig}{\begin{figure}}
\newcommand{\efig}{\end{figure}}

\usepackage{hyperref}
\hypersetup{
    colorlinks=true,      
    urlcolor=blue,
    citecolor=blue,
    linkcolor=blue
}

\begin{document}
\title{Dynamics of interacting particles on a rhombus chain: Aharonov-Bohm caging and inverse Anderson transition}

\author{Sitaram Maity$^{1,2}$}
\thanks{These authors contributed equally to this work}
\author{Biswajit Paul$^{1,2}$}
\thanks{These authors contributed equally to this work}
\author{Soumya Prakash Sharma$^{1,2}$} 
\author{Tapan Mishra$^{1,2}$}
\email{mishratapan@niser.ac.in}
\affiliation{
$^1$School of Physical Sciences, National Institute of Science Education and Research, Jatni 752050, India\\
$^2$Homi Bhabha National Institute, Training School Complex, Anushaktinagar, Mumbai 400094, India}

\date{\today}

\begin{abstract}
The Aharonov-Bohm (AB) caging is the phenomenon of extreme localization of particles experiencing magnetic field in certain tight binding lattices. While the AB caging involves the localization of non-interacting particles, it often breaks down due to the effect of interaction resulting in delocalization. In this study, however, we show that interactions under proper conditions can restore the AB caging of particles. By analysing the dynamics of two bosons possessing both onsite and nearest neighbor interactions on a one dimensional diamond/rhombus lattice pierced by an artificial gauge field, we show that the AB caging is restored when both the interactions are of equal strengths. Furthermore, the AB caged bosons, with the onset of an antisymmetric correlated onsite disorder in the lattice, escape from the cages, demonstrating the phenomenon of inverse Anderson transition which is known to be exhibited by the non-interacting AB caged particles. We also obtain situation similar to the inverse Anderson transition when an external potential gradient is applied to the lattice. These findings offer route to realize the AB caging and inverse Anderson transition of interacting particles in experiments involving ultracold atoms in optical lattices or superconducting circuits. 

\end{abstract}

\maketitle

\section{Introduction}

Localization of quantum states and associated transport in lattice systems have been the topic of great interest due to their fundamental relevance and possible technological application. Starting from the seminal work of Anderson involving the localization of electronic states in disorder systems~\cite{Anderson_loc, Abrahams1979,Lee1985}, localization transitions have been predicted and observed in numerous physical systems enhancing our understanding about this exotic phenomenon of nature~\cite{ Roati2008,Maciá2014,Michael2015,uschen2018,Castro2019}. 
While traditionally, the localization transitions are the effect of disorder in the system, certain types of localization occurs due to lattice topology~\cite{Sutherland1986}. One such case of the latter type is the Aharonov-Bohm (AB) caging~\cite{Vidal1998} which is a fascinating phenomenon of flat-band localization of particles in certain lattice systems~\cite{Dou2002, Takayoshi2013, Tovmasyan2013, Flach_2014, Tovmasyan2016}. In particular, the diamond lattice (also known as the rhombus lattice), when subjected to external magnetic fields the destructive interference of tunneling pathways leads to the formation of perfectly flat energy bands, transforming the delocalized Bloch waves into compact localized states (CLSs) that are confined within the individual unit cells of the lattice. This intriguing localization mechanism has been realized experimentally in photonic and atomic systems providing platforms for deeper understanding of the caging mechanism~\cite{Mukherjee2018, Liberto2019, Kremer2020, Longhi_prl2022, Tao2024}.

Recent studies have shown that external perturbations such as interaction and disorder are known to have serious impact on the stability of the AB caging mechanism.
While on one hand, the increasing disorder tends to delocalize the CLSs - a phenomenon known as the inverse Anderson transition (IAT)~\cite{Goda2006}, on the other hand, interaction effects introduce further complexities due to the overlap between the localized single particle states resulting  in the delocalization of the CLSs and hence the breakdown of the AB caging of the interacting particles~\cite{Vidal2000, Jeronimo2023, chen2024}. 
Various studies have been performed to explore and understand the breakdown and possible stability of the AB caging in the presence of interactions~\cite{Rizzi2017, paivi_torma2018, Danieli2020, Roy2020, Auditya2023}. Recent studies have shown that specific interaction couplings can  favour flatband scenarios in the many-body systems causing compact localization of the many-body states dubbed as the many-body flatband localization~\cite{Danieli2020, Nicolau2023}. 
On the other hand, besides strong correlation, effect of classical non-linearity has been explored in the context of weakly interacting Bose systems where the particles are found to undergo a breathing motion~\cite{Liberto2019,Danieli_non-linear-caging2021,Gligori_non-linear2019}.  
Recently, efforts have been made to observe interaction effects on the AB caging of the particles in experiments. In this context, quench dynamics of a pair of interacting particles also known as the quantum walk have been experimentally studied to gain insights about the fate of the AB caging due to  interactions. It has already been shown that the AB caging of two particles is destroyed when local interaction is considered in an experiment involving superconducting circuits~\cite{Jeronimo2023}. On the other hand, experiment using Rydberg excited atoms has revealed that the AB caging of two particles with non-local interaction is also destroyed~\cite{chen2024}. However, the combined effect of both local and non-local interactions on the two particle dynamics is still not well explored. As it has already been demonstrated that competing interactions in the quantum walk reveals novel scenarios~\cite{Longhi_BZO, Mondal2020, giri2022, paul2024}, it will be interesting to study the effect of such competing interactions on the AB caging of interacting particles

In this work, we study the quantum walk of two bosons possessing   both onsite and nearest-neighbour (NN) interactions on a rhombous chain pierced by an artificial gauge field of $\pi$- flux strength (or half a flux quantum per plaquette) and show that when the strengths of the two competing interactions are equal to each other then the two particle states exhibit compact localization and hence an AB caging scenario.  Remarkably, the interaction induced AB caging of two particles also supports the phenomenon of inverse Anderson transition (IAT) when subjected to an onsite lattice disorder - a phenomenon reminiscent of the IAT of the non-interacting particles. We also obtain signatures similar to that of IAT in the presence of external potential gradient or tilt. 

\begin{figure}[t]
    \centering
    \includegraphics[width=1\columnwidth]{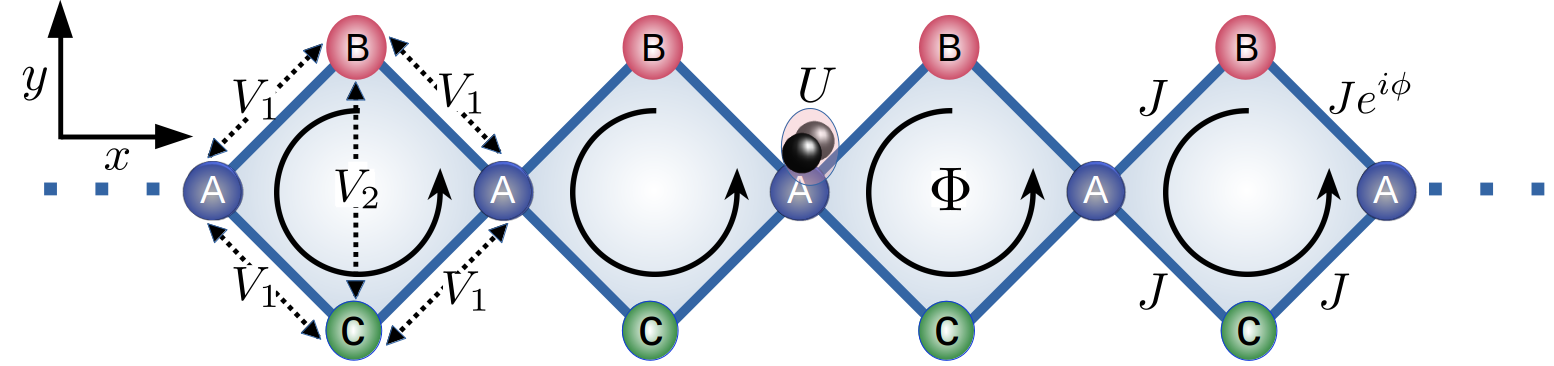}
    \caption{Illustration of a diamond lattice structure with interactions and uniform flux. The unit cell consists of 3 lattice sites such as A, B, and C. Solid lines indicate allowed hopping paths for bosons, with hopping strengths represented by $J$. $U$ denotes the strength of onsite interaction between a pair of bosons. While $V_1$ represents the NN interaction along the sides, $V_2$ represents the interaction along the vertical diagonal of the rhombus. Each plaquette is pierced by a uniform magnetic flux $\Phi$. }
    \label{fig:1}
\end{figure}

This paper is structured as follows: In Section II, we present the model for the one dimensional rhombus lattice subjected to the uniform magnetic flux generated by an artificial gauge field for bosons. In Section III, we provide the main results, i.e., restoration of the AB caging with tunable interactions and the  IAT in an interacting system. Lastly, in Section IV, we provide a brief summary of our results.

\section{Model}
We consider a system of bosons possessing both onsite and NN interactions and occupying the sites of a diamond lattice chain or the chain of stacked rhombi through their vertices in one direction pierced by an artificial gauge field as represented in Fig.~\ref{fig:1}. Here, each vertex represents a lattice site. 
The system under consideration is described by the extended Bose-Hubbard model Hamiltonian with magnetic flux given as 
\begin{equation}
\begin{split}
\hat{H} &= -J \sum_{j} ( \hat{b}_{A,j}^{\dagger} \hat{b}_{B,j} + \hat{b}_{A,j}^{\dagger} \hat{b}_{C,j} + \hat{b}_{A,j+1}^{\dagger} \hat{b}_{C,j} \\
&\quad
+ e^{i\phi}  \hat{b}_{A,j+1}^{\dagger} \hat{b}_{B,j} + \text{H.c.} )  + \frac{U}{2} \sum_{j,\sigma } \hat{n}_{\sigma,j} (\hat{n}_{\sigma,j} - 1) \\
&\quad + V_1 \sum_j ( 
 \hat{n}_{A,j} \hat{n}_{B,j}  + \hat{n}_{A,j} \hat{n}_{C,j} + \hat{n}_{A,j+1} \hat{n}_{B,j} \\
 &\quad +
 \hat{n}_{A,j+1} \hat{n}_{C,j} )
  + V_2 \sum_j  
 \hat{n}_{B,j} \hat{n}_{C,j}.
 \label{eq:ham}
\end{split}
\end{equation}
Here, $\hat{b}_{\sigma,j}^{\dagger} (\hat{b}_{\sigma,j})$ is the bosonic creation (annihilation) operator at $\sigma \in (A,B,C)$ lattice site of the $j^{th}$ unit cell and $\hat{n}_{\sigma,j} = \hat{b}_{\sigma,j}^{\dagger} \hat{b}_{\sigma,j}$ is the number operator at the $\sigma$ site of the $j^{th}$ unit cell. $J$ represents the hopping strength between the NN sites and  $U$ denotes the strength of the on-site interaction. $V_1$ and $V_2$ are the NN interaction strengths along the sides and the vertical diagonal of the rhombus, respectively. This means we assume the NN interactions between the A-B, A-C, and B-C sites only. 
Here, $\phi$ is the Peierl's phase~\cite{Peierls1933} acquired by the particle due to an uniform flux piercing through each plaquette of the lattice.

We analyse the  dynamics by following the standard protocol given as 
\begin{equation}
   |\psi(t)\rangle=e^{-i\hat{H}t}|\psi(0)\rangle 
\end{equation}
by solving the time-dependent Schr\"{o}dinger equation, where $\hat{H}$ is the Hamiltonian shown in Eq.~\ref{eq:ham}. Here, $|\psi(0)\rangle$ is the initial state. We numerically investigate the system by implementing the exact diagonalization(ED) method for sufficiently large systems. Throughout the work, we fix $\phi =\pi$  which turns all bands flat in the non-interacting limit. We fix $J=1$ which also sets the energy scales and assume $V_1=V_2=V$ in the system. 
In the following, we will discuss the restoration of the AB caging  and inverse Anderson transition obtained in this system.

\section{Results}
In this section, we discuss the dynamics of two interacting particles on the rhombus chain. In the first part, we show how competing interaction can restore the AB caging of the particles and in the second part we focus on the inverse Anderson transition. 

\subsection{Restoration of AB caging}  
It has been well understood that inter-particle interaction turns the flatbands of the $\pi$-flux rhombus lattice dispersive. As a result of this, the CLSs become delocalized and the AB caging is destroyed. 
In this section, we will discuss how competing onsite  and NN interactions among the bosons favors CLSs and hence restore the AB caging in the system. To investigate such a phenomenon we study the dynamics of two interacting bosons  starting from the initial state,
\begin{equation}
    |\psi(0)\rangle = (\hat{b}_{A,j}^\dagger)^2 |0\rangle
    \label{eq:ini1}
\end{equation}
and allowing finite onsite repulsion ($U>J$) and vanishing NN interaction ($V=0$). Such a situation guarantees that for a $\pi$-flux rhombus lattice the AB-caging is already broken due to strong onsite repulsion $U$. With this setup in hand we perform an interaction quench for $V$ which we will show leads to the restoration of the AB caging under proper condition.
\begin{figure}[t!]
    \centering
   \includegraphics[width=1\columnwidth]{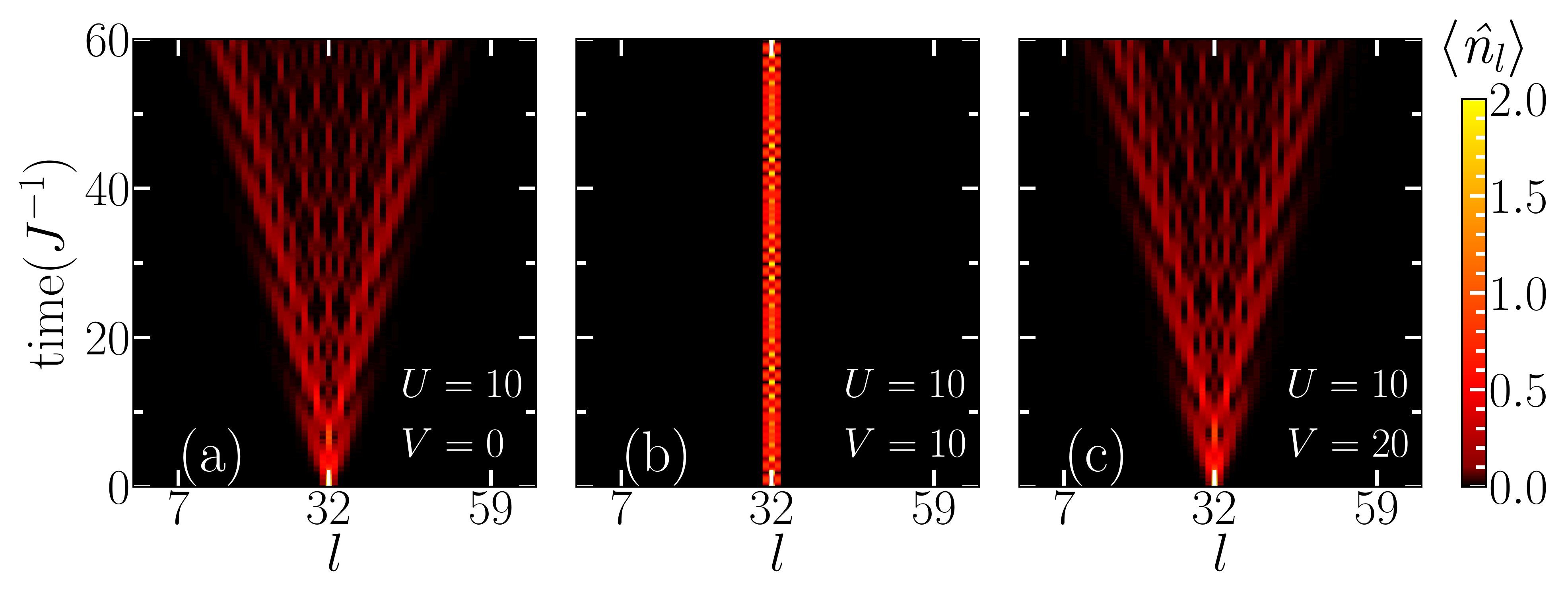}
    \caption{Density evolution is plotted against modified index ($l$) for an initial state $|\psi(0)\rangle = (\hat{b}_{A,16}^\dagger)^2 |0\rangle$ with onsite interactions   $U=10$, and NN interaction  (a) $V=0$, (b) $V=10$, (c) $V=20$ for $L=100$. }
    \label{fig:density_evolution_onsite}
\end{figure}

To understand the dynamics we first analyse the time evolution of real space onsite particle density 
\begin{equation}
    \langle \hat{n}_{ \sigma,j}(t) \rangle=\langle \psi(t)|\hat{b}_{\sigma, j}^\dagger \hat{b}_{\sigma, j}|\psi(t)\rangle
\end{equation}
which provides a first hand picture about the states of the system. Due to the rhombic geometry, we define a new index $l$ in such a way that for all the A-sites, $l=2j$ and for B and C sites together,  $l=2j+1$, where $j$ is the unit-cell index. According to the above notation we redefine the average density in terms of the modified index ($l$) as,
\begin{equation}
\begin{split}
    &\langle \hat{n}_{2j}(t)\rangle = \langle \hat{n}_{A,j}(t)\rangle \ \ \text{and} \\  &\langle \hat{n}_{2j+1}(t)\rangle = \langle \hat{n}_{B,j}(t)\rangle+\langle \hat{n}_{C,j}(t)\rangle
    \label{eq:density}
\end{split}
\end{equation}
and  $j$ starts from $0$ in our case. We plot the average densities $\langle n_l\rangle$ in Fig.~\ref{fig:density_evolution_onsite} for $U=10$ and for different values of $V$.

In the presence of strong onsite interaction $U=10$ and vanishing NN interaction $V=0$ and for the choice of the initial state, the two-particles form a repulsively bound pair~\cite{Winkler2006} of bosons, which behaves effectively as a single particle with reduced hopping strength $J_{eff} = \frac{J^2}{U}$. This bound pair experiences an effective flux $\phi = 2\pi$ while the actual flux threaded through each plaquette is $\phi=\pi$. Therefore, the bound pair which behaves like an effective single particle does not experience any effect from the external gauge field. Such a situation is not conducive for the AB caging and the particles smoothly escape the cage leading to delocalization of states~\cite{Jeronimo2023}. In such a scenario we obtain a light-cone type spreading of the onsite particle density as shown in Fig.~\ref{fig:density_evolution_onsite}(a).  Surprisingly, when the NN interaction strength $V$ is finite and equal to the onsite interaction strength $U$ (i.e. $V=U=10$ in this case), we obtain that the spreading of the onsite density is completely suppressed and the particles remain localized within a narrow region around the initial position in the long time dynamics as shown in Fig.~\ref{fig:density_evolution_onsite}(b). This behavior demonstrates the signature of compact localization of states and the restoration of the AB caging of particles. 
In contrast, when $V>U$ the particles exhibit a tendency to spread across the entire lattice which is shown in Fig.~\ref{fig:density_evolution_onsite}(c) for $V=20$. The evolution of the onsite particle density shows significant redistribution, leading to a more uniform density profile over time. This indicates that the particles explore the lattice extensively, resulting in a broader distribution of density across the sites. The spreading of the particles from their initially confined site is an indication of the absence of AB caging under these conditions. 

\begin{figure}[!t]
    \centering
\includegraphics[width=0.51\textwidth]{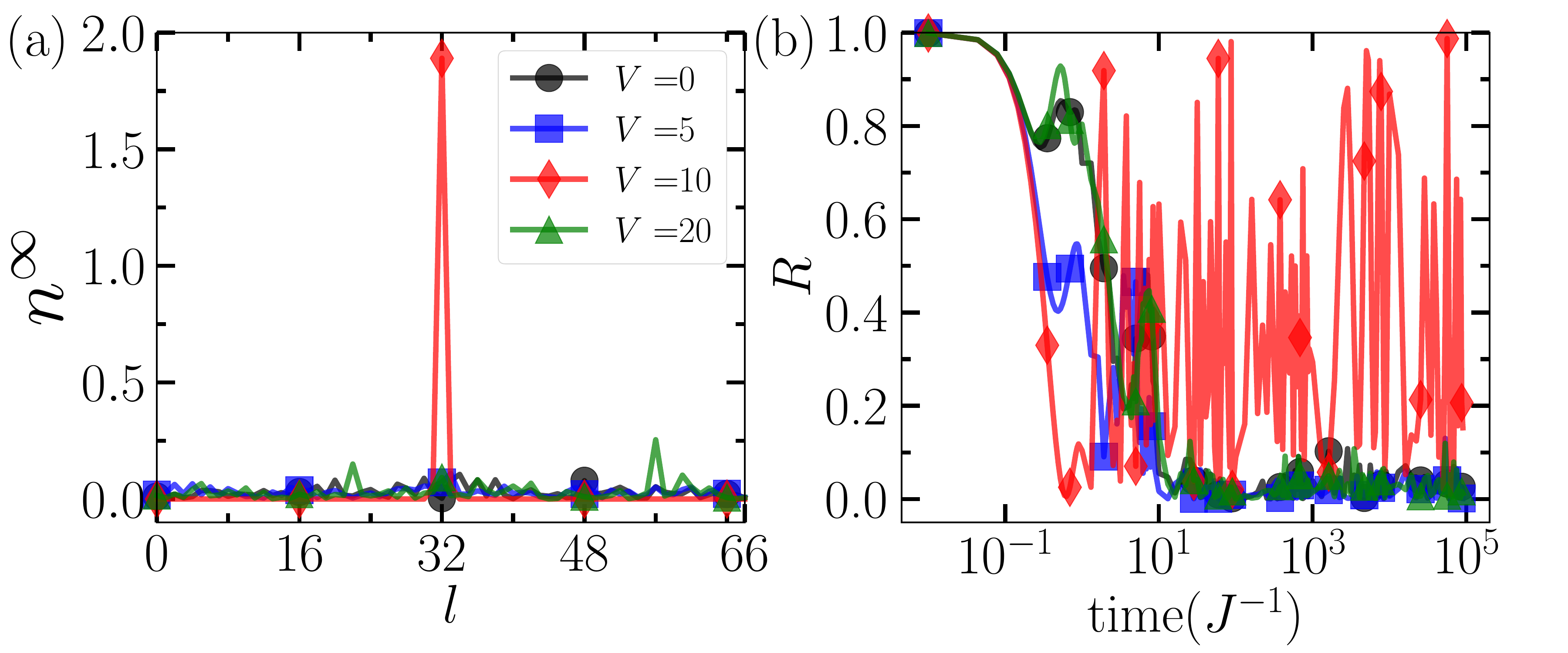}
    \caption{(a) $n^\infty$ plotted against $l$ after a sufficiently long time  ($t\sim 10^8(J^{-1})$)  for different NN interaction strengths $V=0,~5,~10$ and $~20$. (b) Return probability ($R$) as a function of $t(J^{-1})$ for different values of $V$. In both the figures, we consider the initial state $|\psi(0)\rangle = (\hat{b}_{A,16}^\dagger)^2 |0\rangle$ and $U=10$ for $L=100$ sites.}
    \label{fig:return_prob_onsite}
\end{figure}

To obtain further inference from the density evolution, we plot the densities after a very long-time evolution (e.g. at $t \sim 10^8 (J^{-1})$) defined as 
\begin{equation}
    {n}^\infty=\langle \psi(t\to\infty)|\hat{b}_{\sigma, j}^\dagger \hat{b}_{\sigma, j}|\psi(t\to\infty)\rangle
\end{equation}
against the modified index $l$ in Fig.~\ref{fig:return_prob_onsite}(a) which is calculated in a similar way as defined in Eq.~\ref{eq:density}. The localization of particles near the initial position for $V=10$ compared to other values of $V$ can be seen as a sharp peak in $n^\infty$ (red diamonds).
The behavior shown in Fig.~\ref{fig:density_evolution_onsite}(b) together with Fig.~\ref{fig:return_prob_onsite}(a) suggests that the combined effect of $U$ and $V$ favors localization of particles and hence the restoration of the AB caging at $U=V=10$.

To confirm the restoration of the AB caging of particles, we analyse the time evolution of the return probability, $R$, which measures the likelihood of finding the particle back at its initial position after a given time $t (J^{-1})$ and is defined as,
\begin{equation}
    R(t) = |\langle \psi(0) | \psi(t) \rangle|^2
\label{eq:return_prob}
\end{equation} 
In Fig.~\ref{fig:return_prob_onsite}(b), we plot $R$ as a function of $t(J^{-1})$ for values of $V=0,~5,~10$ and $20$. It can be seen that $R$ saturates to zero in the long time limit for all values of $V$ except for $V=10$ (red diamonds) where it exhibits an oscillatory behaviour indicating a high likelihood that the particles remain localized at their initial positions. The high values of $R$ for $U=V=10$ is the signature of the AB caging. Conversely, the vanishing values of $R$ for $U\neq V$  reflect the extensive spreading of particles across the lattice, as they are less likely to return to their initial positions. This behaviour is also consistent with the uniform density profiles observed for $U\neq V$. 

\begin{figure}
    \centering
    \includegraphics[width=1\columnwidth]{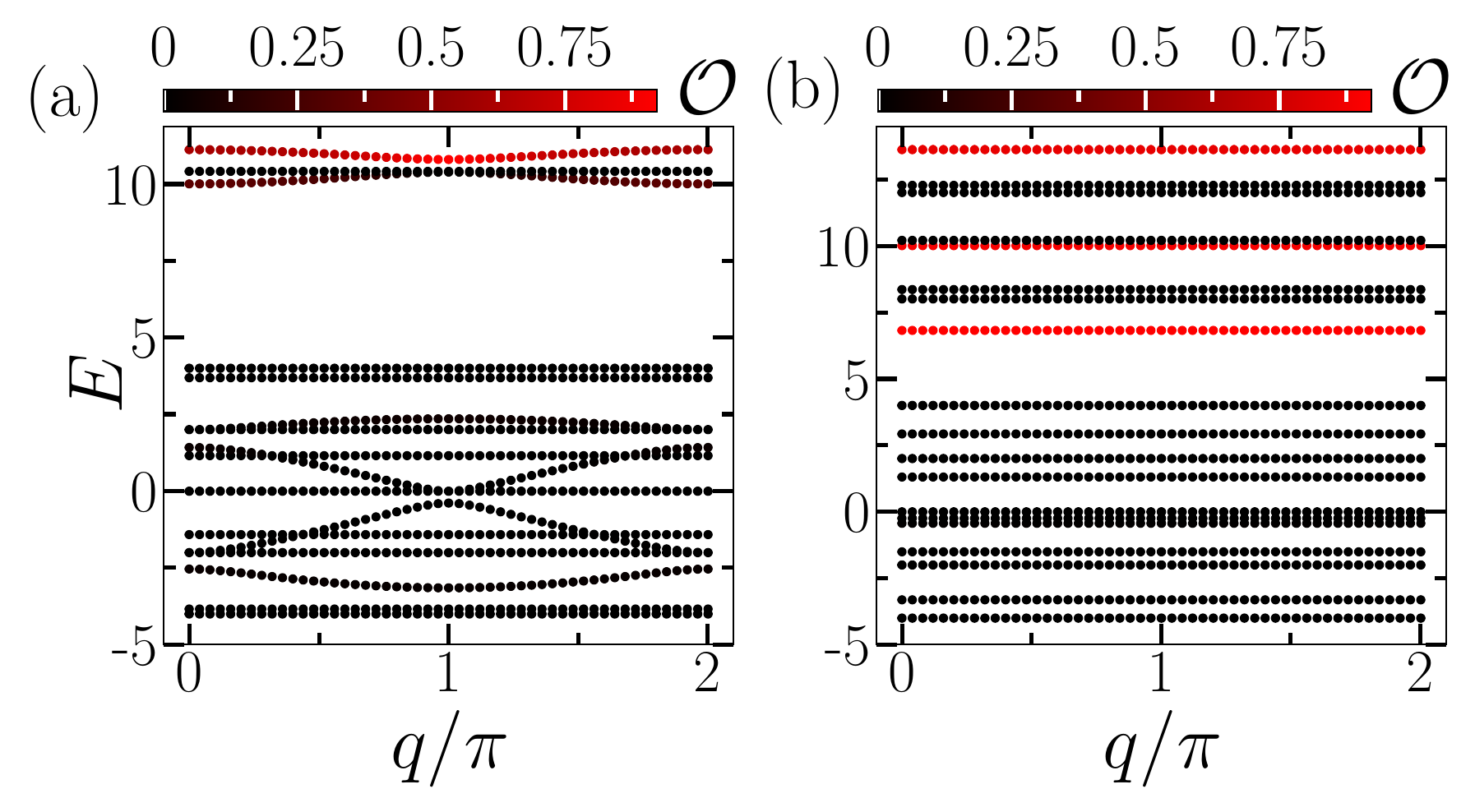}
    \caption{Band structure for two interacting bosons with system size $L=150$ with periodic boundary condition for  $U=10$, $\phi=\pi$ and NN interactions (a) $V=0$ and (b) $V=10$. The colors of individual eigenstates indicate the overlap ($\mathcal{O}$) with the initial state.}
    \label{fig:band}
\end{figure}

To delve deeper into the restoration of the AB caging phenomenon, we examine the band structure corresponding the model under consideration for different values of $V$ while keeping $U$ fixed. We plot the band structure by numerically diagonalizing the Hamiltonian in momentum space. We also compute the overlap
\begin{equation}
\mathcal{O}=|\langle\Psi|\chi_i\rangle|^2,
\end{equation}
where, $|\Psi\rangle$ is the state in the momentum space corresponding to the initial state given in Eq.~\ref{eq:ini1} and $|\chi_i\rangle$ are the momentum space eigenstates of the Hamiltonian in Eq.~\ref{eq:ham} obtained using periodic boundary condition. The finite values of $\mathcal{O}$ provides information about the states which participate in the dynamics.
In Fig.~\ref{fig:band} we plot the band structure obtained for a periodic system of size $L=150$ together with $\mathcal{O}$. 
From Fig.~\ref{fig:band}(a), we observe that for $U=10$ and $V=0$, the bands with finite values of $\mathcal{O}$  are dispersive which is indicative of delocalized states. This dispersion suggests that the particles are not confined to any specific regions of the lattice and no AB caging phenomenon.
However, upon increasing the values of $V$ to $U = V = 10$, we observe the emergence of flat bands in the energy spectrum, as shown in Fig.~\ref{fig:band}(b). These flat bands are responsible for the localization of particles and hence the restoration of the AB caging. 
From this analysis, it is evident that only the bands with higher energies play significant roles in determining the dynamics of the system and the underlying mechanism of the AB caging. The equal strengths of the onsite and NN interaction favor the regeneration of the flatbands in the system due to which the AB caging of interacting particles is restored. 

\begin{figure}[t!]
    \centering
   \includegraphics[width=1\columnwidth]{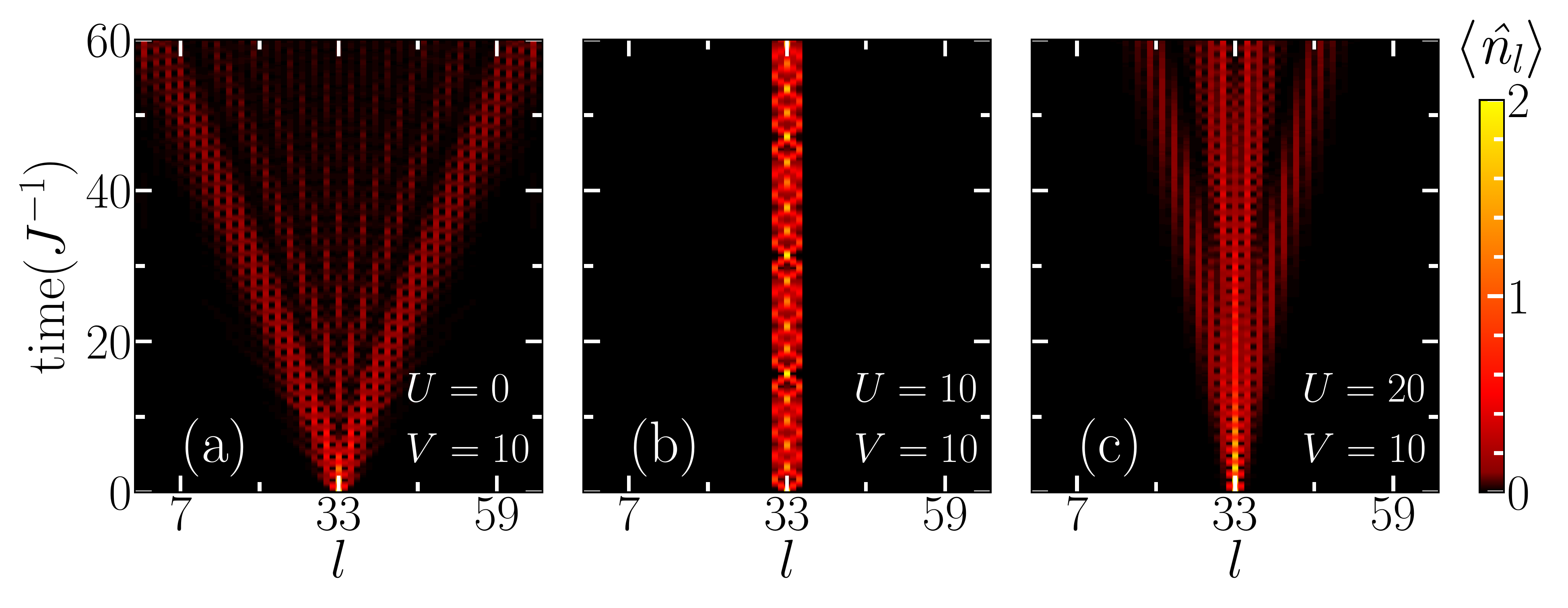}
    \caption{Density evolution is plotted against modified index ($l$), starting from an initial state $|\psi(0)\rangle = \hat{b}_{B,16}^\dagger \hat{b}_{C,16}^\dagger |0\rangle$ for $V=10$, and onsite interaction (a) $U=0$, (b) $U=10$, (c) $U=20$. Here the system of $L=100$ is considered.}
    \label{fig:density_evolution_rung}
\end{figure}

To understand the dependence of the AB caging on the choice of initial state,  we consider a two particle initial state by allowing one particle each at sites B and C of the same unit-cell  which is given as,
\begin{equation}
    |\psi(0)\rangle = \hat{b}_{B,j}^\dagger \hat{b}_{C,j}^\dagger |0\rangle.
    \label{eq:ini2}
\end{equation}
In this case, contrast to the previous case we fix $V = 10$ at the initial time and vary the onsite interaction $U$ to investigate the dynamics. In the absence of $U$, the onsite densities of particles exhibits a linear spread indicating delocalization as shown in  Fig.~\ref{fig:density_evolution_rung}(a). With increase in $U$ we obtain similar behaviour obtained for the previous case i.e.  when $V=U=10$, the dynamics exhibits signatures of the AB caging phenomenon  as shown in Fig.~\ref{fig:density_evolution_rung}(b) (compare with Fig.~\ref{fig:density_evolution_onsite}(b)). Also in the limit when $U>V$, the AB caging is broken (see Fig.~\ref{fig:density_evolution_rung}(c)). However, in this case, the extent of the cage is larger compared to the situation when the two particles are initially at site A (compare Fig.~\ref{fig:density_evolution_onsite}(b) and Fig.~\ref{fig:density_evolution_rung}(b)). Note that we also obtain the restoration of the AB caging for any two-particle initial state that corresponds to the two particles residing within the unit cell. 

All the above signatures reveal that the AB caging in the two particles' dynamics is restored when the onsite and the NN interactions are of equal strengths ($U=V$).  It is to be noted that it is sufficient to establish the restoration of the AB caging of two interacting particles with  the onsite interaction $U$ and the NN interaction only along the BC bond i.e. $V_2$. The finite values of $V_1$ do not affect the caging phenomenon as long as they are of equal strengths. For unequal values of $V_1$ across the four bonds results in the breakdown of the AB caging (not shown).

The above analysis shows a clear signature of the restoration of the AB caging for two interacting particles on a $\pi$-flux rhombus chain.
At this point, it is natural to explore the fate of such interaction-induced AB caging in the presence of external perturbations which we will discuss in the rest of the paper.

\subsection{Inverse Anderson Transition}
One of the remarkable manifestation of the AB caged particles is inverse Anderson transition (IAT) which is a phenomenon of non-trivial delocalization of CLSs due to the effect of disorder. In recent years, the IAT has been widely studied in various platforms both theoretically and experimentally ~\cite{Longhi2021, Longhi_prl2022, Wang2022, Wang_non_abelian_IAT2023}.
While the IAT is known to be exhibited by the AB caged non-interacting particles, in this case we will show that the AB caged interacting particles also exhibit IAT when subjected to onsite disorder.

\begin{figure}[t!]
    \centering
    \includegraphics[width=0.49\textwidth]{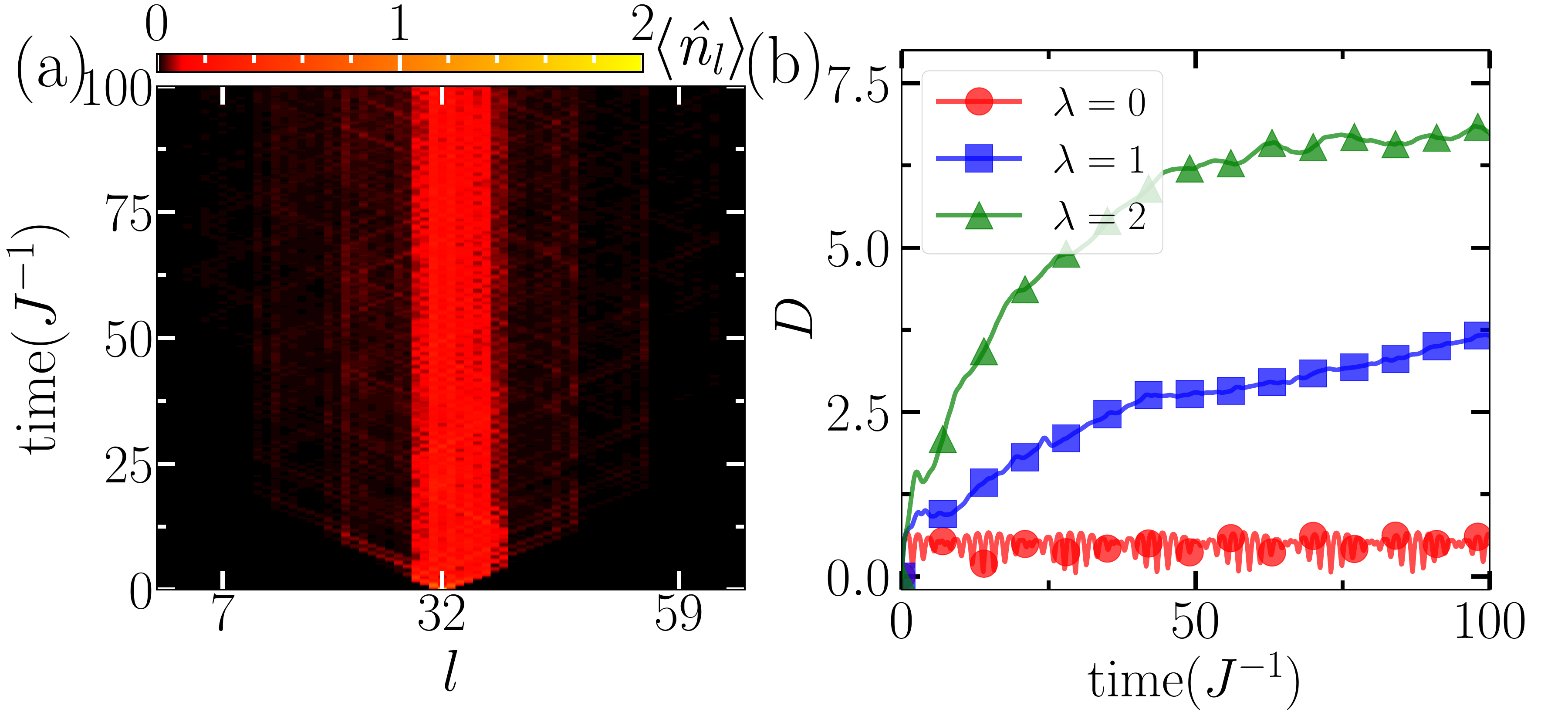}
    \caption{(a) Density evolution with antisymmetric quasiperiodic disorder is considered. The disorder strength we consider in this case is $\lambda = 2$. (b) Shows the root mean square displacement ($D$) as a function of time for antisymmetric quasiperiodic disorder strengths $\lambda=0$ (red circles), $1$ (blue squares), $2$ (green triangles). In all of these above cases, we consider $U=V_2=10$ for $L=100$.}
    \label{fig:quasi_den+roe}
\end{figure}

To this end we introduce the disorder term in the Hamiltonian given in Eq.~\ref{eq:ham} as
\begin{equation}
    \hat{H'}= \hat{H}+\hat{H}_{\text{dis}},
\end{equation}
where $  \hat{H}_{\text{dis}} = \sum_{\sigma, j} \lambda_{\sigma, j} \hat{n}_{\sigma, j}$ and $\lambda_{\sigma, j}$ is the onsite disorder potential at unit-cell $j$ and site $\sigma$. In our case, we consider quasi-periodic disorder~\cite{aubry1980analyticity, Roati2008, Lahini2009, Biddle2010, Biddle2011, Henrik2018, hobbyhorse_2019, Li2020, shilpi2021, Padhan2022, Vu2023} and note that if random disorder is considered in place of quasiperiodic one the result obtained is qualitatively similar. To achieve IAT we choose $\lambda_{\sigma, j}=\lambda \cos(2\pi \beta j + \Delta)$, where $\lambda$ is the disorder strength and $\beta$ is an irrational number (typically chosen as the Golden ratio $\beta = \frac{\sqrt{5} - 1}{2}$), and $\Delta$ is a random phase offset. 
Disorder average is taken over $500$ offset values. 
We choose two types of quasiperiodic disorder in our model: symmetric and antisymmetric correlated disorder~\cite{Longhi_prl2022, Longhi2021, Ahmed2022}. In the symmetric case, the strength of quasi-periodic disorder potential is set identical for both lattice sites B and C i.e., $\lambda_{B, j}=\lambda_{C, j}$, while in the antisymmetric case, $\lambda_{B, j} = - \lambda_{C, j}$ is chosen. In both cases we set $\lambda_{A,j}=0$. We obtain that the symmetric choice of the quasiperiodic disorder does not lead to IAT. However, if antisymmetric disorder is considered the system undergoes an IAT as in the case of non-interacting particles. 

\begin{figure}[t!]
    \centering
    \includegraphics[width=0.49\textwidth]{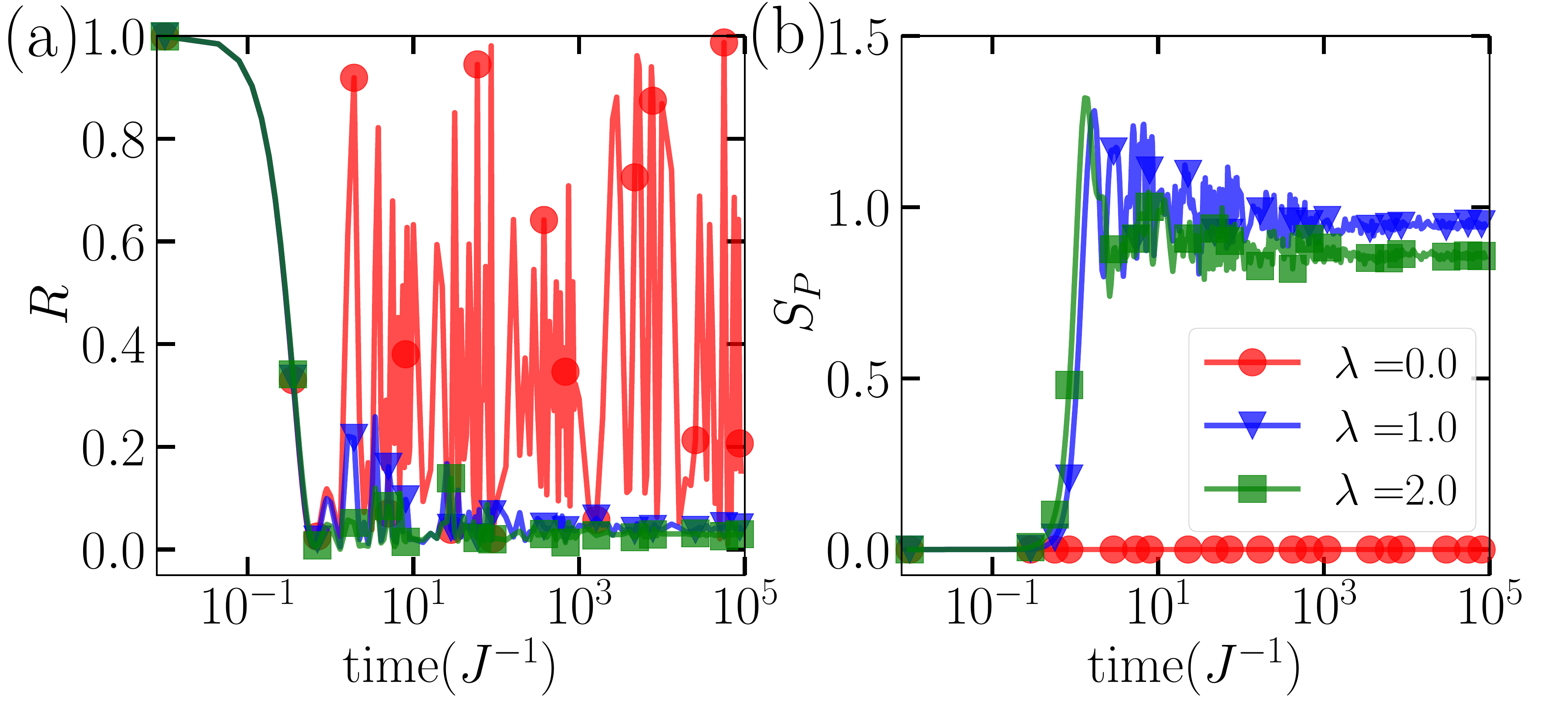}
    \caption{(a) and (b) show the return probability ($R$), and bipartite EE ($S_P$) as a function of time for correlated antisymmetric quasiperiodic disorder strengths $\lambda=0$ (red circles), $1$ (blue triangles), $2$ (green squares). In all of these above cases, we consider $U=V=10$, and (a) and (b) are plotted for system size $L=100$ and $L=49$, respectively.}
    \label{fig:quasi_Rt+entropy}
\end{figure}

We start by fixing $U=V=10$ for which the particles are already AB caged and then study the dynamics by varying the disorder strength $\lambda$. First of all, we examine how the time evolution of the onsite densities behave for finite  values of $\lambda$. As depicted in Fig.~\ref{fig:quasi_den+roe}(a), for $\lambda=2$, we obtain a clear spreading of the density over a region of the lattice as compared to no spreading for $\lambda=0$ shown in Fig.~\ref{fig:density_evolution_onsite}(b). Such a behavior in density is due to the escaping of the particles from the AB cage, implying that the initially localized states delocalize. This behavior  can be clearly seen by monitoring the time evolution of the root mean-square displacement defined as 
\begin{equation}
    D(t) = \bigg[\sum_{l} (l-l_0)^2 \langle \hat{n}_l(t) \rangle\bigg]^{1/2}
\label{eq:msd}
\end{equation} 
where $l$ is the modified index and $l_0$ is the initial position of the two particle state. We plot $D$ as a function of $t(J^{-1})$ in Fig.~\ref{fig:quasi_den+roe}(b) for $U=V=10$ for $\lambda=1$ (blue squares) and for $\lambda=2$ (green triangles) and compare it with $\lambda=0$ (red circles). While for $\lambda=0$ we obtain that $D$ immediately saturates to a very small value, for small but finite values of $\lambda$, $D$ increases with time, indicating delocalization of the initially localized states of the AB caged particles. Signatures of this delocalization can also be obtained from the survival probability $R$ that is introduced previously. In Fig.~\ref{fig:quasi_Rt+entropy}(a), we compare the evolution of $R$ for three different values of $\lambda$. The vanishing up of $R$ for $\lambda=1$ (blue triangles) and $\lambda=2$ (green squares) as compared to  $\lambda=0$ (red circles) in the long time dynamics confirms the breakdown of the AB caging and the onset of an IAT. 

Entanglement entropy (EE) is an useful diagnostic to quantify the quantum correlations and localization properties of interacting systems. In the context of the AB caging, EE serves as an indicator of the localization and delocalization of particles. A low EE signifies a high degree of localization, consistent with the AB caging phenomenon. Conversely, an increase in EE indicates delocalization and the breakdown of the AB caging. In our study, we utilize the EE to concretely establish the  IAT or the breakdown of the AB caging due to quasi-disorder. The EE ($S_P$) for a subsystem $P$ is calculated using the reduced density matrix $\rho_P$. For a many-body state $|\psi \rangle$ of the entire system, the reduced density matrix is obtained by tracing out the degrees of freedom of the complementary subsystem $Q$, i.e. $\hat{\rho}_P= Tr_Q (|\psi \rangle \langle \psi| )$. The EE is then given by the von Neumann entropy:
\begin{equation}
    S_P = -Tr(\hat{\rho}_P ~log ~(\hat{\rho}_P))
    \label{eq: entropy}
\end{equation}
For an initial state with particles initialized at the A-site of the k$^{th}$ unit cell, we choose the subsystem $P$ consists of $5$ sites, namely, $B_{k-1}, ~C_{k-1},~A_k,~B_{k+1},~C_{k+1}$. In Fig ~\ref{fig:quasi_Rt+entropy} (b) we plot $S_P$ as a function of $t(J^{-1})$ for different values of $\lambda$.

For the two-particle initial states without any disorder ($\lambda=0$), when $U=V$, we observe zero EE as shown in Fig.~\ref{fig:quasi_Rt+entropy}(b) (red circles), reflecting the AB caging of the particles. Such behaviour in EE is due to the no quantum correlations between the subsystems. However, for finite values of antisymmetric disorder $\lambda=1$ (blue triangles) and $2$ (green squares), $S_P$ rapidly grow and saturate to finite values indicating increased quantum correlations and eventual delocalization of the AB caged particles. This confirms the phenomenon of IAT of the two interacting particles AB caged due to competing onsite and NN interactions.

\begin{figure}[t!]
    \centering
    \includegraphics[width=0.49\textwidth]{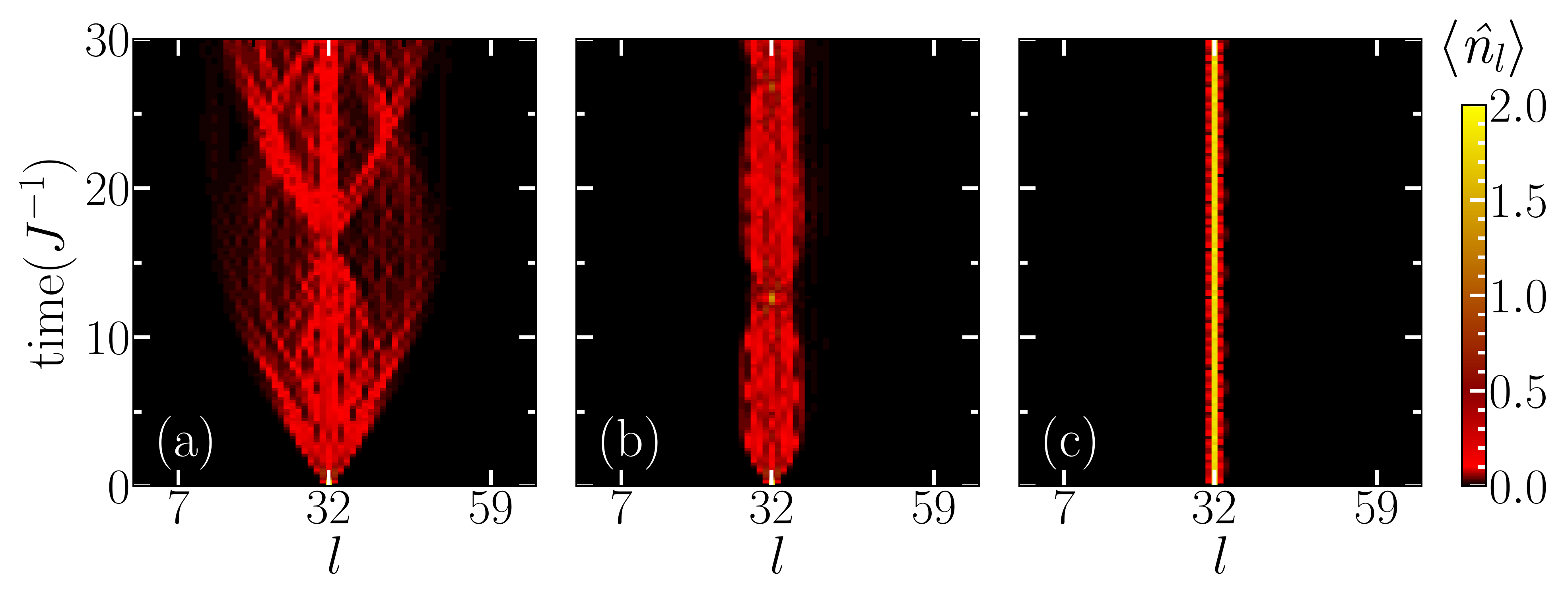}
    \caption{Density evolution is plotted against the modified index ($l$), starting from an initial state $|\psi(0)\rangle = (\hat{b}_{A,16}^\dagger)^2 |0\rangle$ and  $U=10=V=10$ for (a) $F=0.1$, (b) $F=1.0$, (c) $F=10$. The system size considered here is $L=100$.}
    \label{fig:den_tilt}
\end{figure}

Note that by including finite values of $\lambda_{A,j}=\lambda \cos(2\pi \beta j + \delta)$ we also get IAT but eventually the system relocalizes with increasing $\lambda$ (not shown). The asymmetric potential strengths at the B and C sites that favours the IAT of AB caged interacting particles also signals another route to obtain IAT of the two interacting particles  i.e. through external tilt potential which imposes unequal local offsets at each sites of the lattice. In the following we show that a suitable arrangement of tilt also leads to an IAT of interacting particles.

\subsection{IAT due to external tilt or gradient}
To obtain tilt dependent IAT we allow linear potential gradient in both horizontal ($x$) and vertical ($y$) directions such that each sites in the unit-cell experiences different potential strengths. In this case, the Hamiltonian of our system becomes,
\begin{equation}
\begin{split}
    \hat{H}^{\prime} = \hat{H} + \sum_j 2jF \hat{n}_{A, j}&+\sum_j ((2j+1)F+\delta) \hat{n}_{B, j}\\
    &+\sum_j ((2j+1)F-\delta) \hat{n}_{C, j}),   
\end{split}
\end{equation}
where $\hat{H}$ is the Hamiltonian shown in Eq.~\ref{eq:ham}, $j$ is the unit cell index, $F$ is the strength of the potential gradient along the $x$-direction and $\delta$ decides the gradient along the $y$- direction. We study the dynamics starting from the initial state given in Eq.~\ref{eq:ini1}, by maintaining the AB caging condition i.e. $U=V=10$ and change the potential gradient $F$ while fixing $\delta=2$. We obtain that with increase in the values of $F$ the initially localized states start to become delocalized as depicted in Fig~\ref{fig:den_tilt}(a) for $F=0.1$. This delocalization of states due to $F$ is a signature of the IAT due to external potential gradient. With further increase in the value of $F$ the extent of localization start to decrease and eventually for large $F$, the states are localized again which can be seen from Fig~\ref{fig:den_tilt}(b) and (c) plotted for $F=1$ and $F=10$ respectively. 
\begin{figure}[t!]
    \centering
\includegraphics[width=0.49\textwidth]{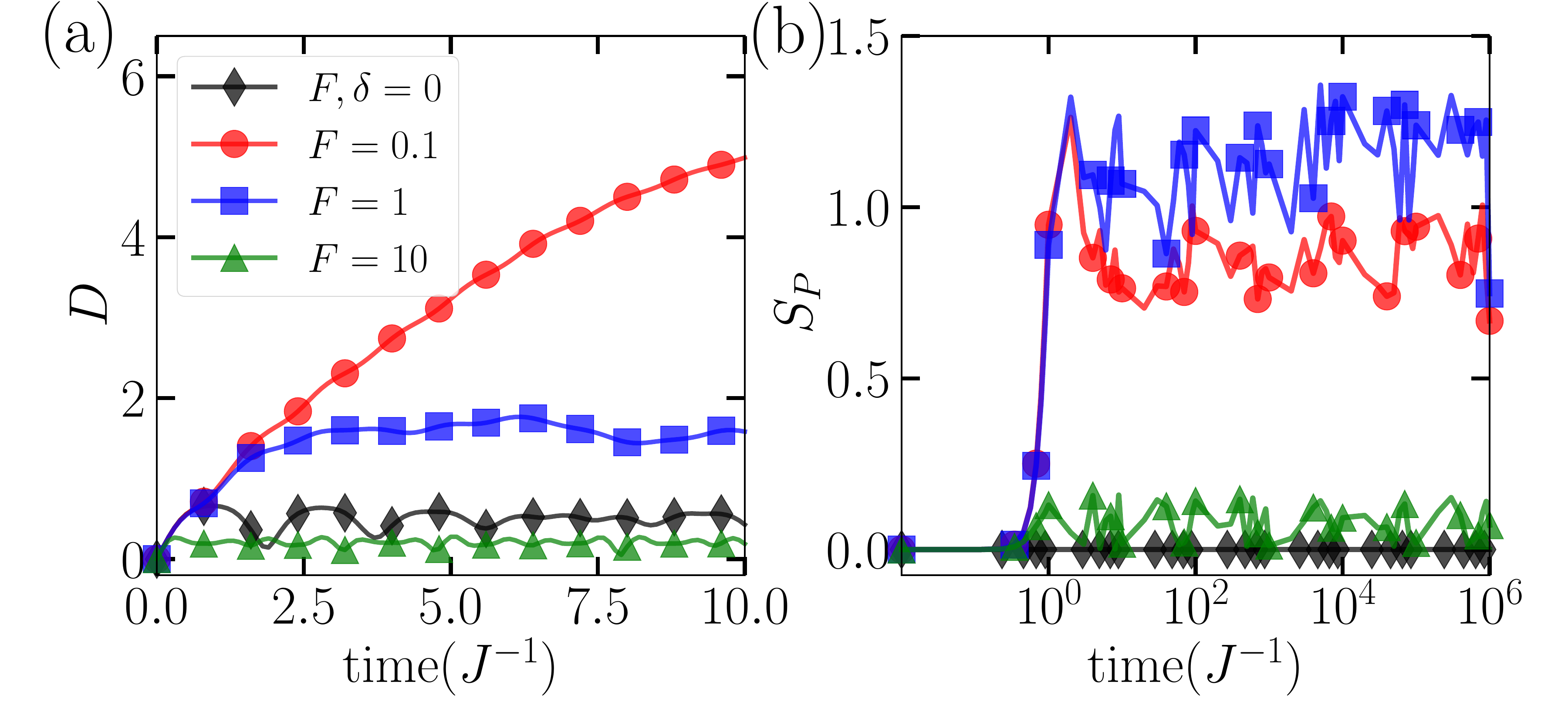}
    \caption{(a) Shows the the root mean-squares displacement ($D$) as a function of $t(J^{-1})$ for the potential gradient $F=0.1$ (red circles), $~1$ (blue squares) and $10$ (green triangles) by keeping interaction strengths $U=V=10$ and fixed $\delta = 2$. The $D$ for $F=\delta=0$ (black diamonds) is shown for comparison. Here, the system of $L=100$ is considered. (b) Shows the EE  plotted against $t(J^{-1})$ for similar parameters as in (a) for $L=49$.}
    \label{fig:tilt_ROE+entropy}
\end{figure}
For further quantification of IAT of interacting particles through potential gradient, we compute the radius of expansion $D$ and time evolution of entanglement entropy $S_P$ defined before. In Fig.~\ref{fig:tilt_ROE+entropy}(a) and (b) we plot $D$ and $S_P$ respectively, as a function of $t(J^{-1})$ for different values of $F$. The increase in $D$ with time in Fig.~\ref{fig:tilt_ROE+entropy}(a) for $F=0.1$  (red circles) indicates the spreading of particle in the lattice compared to that for the lattice without any tilt ($F,~\delta=0$) for which $D$ saturates to a value close to zero (black diamonds) due to the AB caging. Such behaviour in $D$ is due to the IAT in the system. However, for $F=1$ (blue squares), $D$ tends to saturate to a finite value and with increase in $F$, the value at which $D$ saturates to, decreases further as can be seen for $F=10$ ( green triangles). 


Similar features are also seen in the case of $S_P$ which is shown in Fig~\ref{fig:tilt_ROE+entropy}(b) for the values of $F$ considered in Fig~\ref{fig:tilt_ROE+entropy}(a). For the calculation of entropy we have chosen the same bi-partition of the system as is done in the previous section . For small or intermediate values of $F$ (i.e. for $F=0.1$ and $1$) the entanglement entropy rapidly grows and saturates to finite values indicating the breaking of the AB caging. However, for $F=10$, $S_P$ saturates to a very small value due to localization mediated by the potential gradient. For comparison we also show $S_P$ as a function of $t(J^{-1})$ for $F=\delta=0$ for which we have the AB caging. Note that the eventual localization as a function of $F$ is due to the Stark localization of the interacting particles~\cite{Stark_MBL2019}.

\section{Conclusion}
In this work, we have studied the dynamics of two interacting bosons on a $\pi$-flux rhombus chain which is one of the standard models to study physics due to flatbands. While the recent experimental studies based on dynamic of two interacting particles on rhombus chains have revealed that allowing  inter-particle onsite or NN interaction leads to the breakdown of the AB caging of the non-interacting particles, we have obtained that if two competing interactions (onsite and NN interaction) are considered together then the AB caging is favoured in the two-particle dynamics. By analysing the density evolution, return probability and subsystem entanglement entropy we have revealed that the AB caging is restored when the  onsite and the NN interactions are of equal strengths. We have also shown the restoration of the AB caging is independent of the choice of the initial states. In other words, the system exhibits AB caging irrespective of where the two particles are initialized within a particular unit-cell. Such interaction induced restoration of the AB caging is found to occur even for weak interaction strengths. Most importantly, we have obtained that the localized states of the AB caged particles become delocalized due to the effect of onsite disorder resulting in the phenomenon of inverse Anderson transition which is a phenomenon exhibited by the non-interacting particles. We have also extended our studies to explore the effect of external gradient or tilt to the lattice and obtained behaviour similar to the IAT of the interacting particles.

Our findings provides a detailed analysis of the dynamics or quantum walk of two interacting particles revealing the restoration of the AB caging on a lattice which is known to exhibit compact localization of the states of the non-interacting particles due to flatband effects. Moreover, our study also provides the effect of external perturbations in the form of quasiperiodic disorder or external tilt leading to the IAT. Such analysis provides a bottom-up approach to understand the dynamics of interacting particles which can be extended to the true many-body limit. Also similar studies can be explored in different lattices possessing flatband effects to understand the competing effects of flatband, interaction, topology, disorder and particle statistics. Most importantly, the dynamics of two interacting bosons in a rhombus lattice  has been experimentally realized using various platforms such as superconducting circuits~\cite{Jeronimo2023}, ultracold atoms in optical lattices~\cite{Longhi_prl2022} and Rydberg excited atoms~\cite{Tao2024, chen2024}. These studies can in principle be extended to incorporate two competing interactions like the one considered here to simulate the restoration of AB caging and IAT of two interacting particles.

\section{Acknowledgement}
T.M. acknowledges support from Science and Engineering Research Board (SERB), Govt. of India, through project No. MTR/2022/000382 and STR/2022/000023.

\bibliography{references}
\end{document}